# Uncertainty amplification due to density/refractive-index gradients in volumetric PTV and BOS experiments

Lalit K. Rajendran[1], Sally P. M. Bane[1] and Pavlos P. Vlachos[2]*

[1]Purdue University, School of Aeronautics and Astronautics, West Lafayette, USA.

[2] Purdue University, School of Mechanical Engineering, West Lafayette, USA.

*pvlachos@purdue.edu

## Abstract

We theoretically analyze the effect of density/refractive-index gradients on the measurement precision of Volumetric Particle Tracking Velocimetry (V-PTV) and Background Oriented Schlieren (BOS) experiments by deriving the Cramer-Rao lower bound (CRLB) for the 2D centroid estimation process. A model is derived for the diffraction limited image of a particle or dot viewed through a medium containing density gradients that includes the effect of various experimental parameters such as the particle depth, viewing direction and f-number. Using the model we show that non-linearities in the density gradient field lead to blurring of the particle/dot image. This blurring amplifies the effect of image noise on the centroid estimation process, leading to an increase in the CRLB and a decrease in the measurement precision. The ratio of position uncertainties of a dot in the reference and gradient images is a function of the ratio of the dot diameters and dot intensities. We term this parameter the Amplification Ratio (AR), and we propose a methodology for estimating the position uncertainties in tracking-based BOS measurements. The theoretical predictions of the particle/dot position estimation variance from the CRLB are compared to ray tracing simulations with good agreement. The uncertainty amplification is also demonstrated on experimental BOS images of flow induced by a spark discharge, where we show that regions of high amplification ratio correspond to regions of density gradients. This analysis elucidates the dependence of the position error on density and refractive-index gradient induced distortion parameters, provides a methodology for accounting its effect on uncertainty quantification and provides a framework for optimizing experiment design.

# Nomenclature

| | | | |
|---|---|---|---|
| $I$ | Image intensity | $K$ | Gladstone-Dale constant |
| $I_{0,r}$ | Peak image intensity for a single light ray | $\rho$ | Density |
| $r$ | Light ray index | $n_0$ | Ambient refractive index |
| $N_R$ | Number of light rays | $x, y, z$ | Coordinates in the object space |
| $X, Y$ | Spatial co-ordinates on the camera sensor/image space | $J_{ij}$ | Fisher Information matrix |
| $\tau$ | Point spread function | $\gamma$ | Gray value per unit exposure |
| $g$ | Gray Level | $d_r$ | Pixel Pitch |
| $k, l$ | Pixel indices | $\hat{n}$ | Thermal Noise |
| $M$ | Magnification | $p$ | Probability Density Function (PDF) |
| $\epsilon$ | Angular Deflection | $f$ | Signal/Measurement |
| $t$ | Spatial co-ordinate on the density field | $\boldsymbol{a}$ | Model Parameter vector |
| $\theta$ | Light Ray angle | $a_i$ | Model Parameter |
| $\eta$ | Standard deviation of the Gaussian intensity profile on the image plane | $\sigma$ | Standard Deviation |
| $Z_D$ | Distance from dot pattern to density gradient field | $\boldsymbol{N}$ | Normal distribution |
| $\delta$ | Dirac delta function | $f_\#$ | F-number |
| $f$ | Signal | $\zeta$ | Particle depth in the density gradient field |
| $\boldsymbol{E}$ | Expectation operator | $m$ | Model |
| $\beta$ | Blurring Coefficient | $u$ | Measurement |
| $\lambda$ | Wavelength of light | $R_{ij}$ | Rotation matrix |
| $\alpha$ | Image exposure | $\Delta\theta_0$ | Angle of the ray cone |
| $AR$ | Amplification Ratio | $\vec{T}$ | Translation vector |

# 1 Introduction

Particle Image Velocimetry (PIV)/Particle Tracking Velocimetry (PTV) and Background Oriented Schlieren (BOS) are widely used flow diagnostic techniques. PIV/PTV are used to measure the local velocity of a fluid flow by tracking the displacement of tracer particles over time. BOS is used to measure density gradients in a fluid by measuring the apparent distortion of a target dot pattern. These techniques have been applied to a wide variety of small and large scale flow fields in both laboratory and industrial facilities [1]–[4].

However, all three techniques suffer from distortions due to density/refractive-index gradients that are present in a wide variety of fluid flow experiments, such as compressible flows, multiphase flows, flows in facilities with curved windows etc. This is because changes in density/refractive index cause refraction of light rays emanating from the light source (particles for PIV/PTV and dots or speckle patterns for BOS). These distortions affect the final image, and since the measurements rely on processing this image, their effect on measurement accuracy and precision is an important concern.

Past work by Elsinga et. al. [5]–[7] analyzed this problem for 2D planar PIV and showed that the effect of aero-optical distortions due to density/refractive-index gradients can be divided into three categories. First, the refraction of light rays emerging from a particle results in an apparent displacement of the particle when viewed on the camera and leads to a bias error in the position. Second, when the particle moves to a different location in the next time instant, the refraction experienced by the light rays, the corresponding apparent displacement of the particle, and the resulting position error will be different, and the difference in the two position errors between the two time instants leads to a bias error in the velocity. Finally, in a situation with very strong spatial variations in the density/refractive index (such as shock waves), the refraction experienced by two light rays emerging from the same particle can be very different, leading to a blurring of the particle image. This blurring is asymmetric and is in the direction of decreasing density. Elsinga et. al. also found that this blur leads to a broadening of the cross-correlation plane leading to increased random error and a decrease in measurement precision [6].

The impact of distortions due to density/refractive-index gradients has not been analyzed for multi-camera volumetric PTV experiments, which is the current state of the art in three-dimensional, three-component (3D3C) velocity measurement [8]–[10]. The additional challenges introduced in these experiments are (1) varying depth of particles within the laser sheet and (2) varying angle between the direction of the density gradient field and the viewing direction for the different cameras. Another challenge is the need to use small apertures (large f-numbers) in order to keep a thick laser in focus, which can reduce the illumination across the field of view and increase the effect of image noise on the measurement. Similarly, these issues are of importance for BOS experiments, where the deflections of the light rays and the apparent displacement of the dots now comprise a signal and not an error, and in these cases, the effect of image blur on the centroid estimation is most important.

For both measurement techniques, tracking-based displacement estimation methods have been shown to be more accurate and precise than correlation-based methods. In volumetric PTV, Lagrangian tracking as proposed in the Shake-The-Box algorithm by Schanz et. al. [10] significantly increases accuracy and spatial resolution, suppresses the identification of ghost particles, and reduces reconstruction time even at high seeding densities, as the temporal information is used to improve particle reconstruction. In BOS, the dot locations on the target are known from the time of manufacture, and this information can be used to improve the spatial resolution and enable the method to be applied even in images with high dot densities [11].

In tracking-based methods, the centroid estimation process from the particle/dot image controls the measurement accuracy and precision. Therefore, the primary effect of distortions due to density/refractive-index gradients is to increase the 2D position error due to the apparent displacement of the particle (bias error) and the particle image blur (random error). In volumetric PTV, this 2D position error from the images will in turn also affect the 3D position error of triangulating a particle's location in the measurement volume. In BOS, the 2D position error increases the noise floor and reduces the dynamic range of the measurement, especially since the displacements in a typical BOS experiment are generally low (< 1 pix.).

We investigate these issues using a theoretical analysis on the effect of density/refractive-index gradients on the measurement precision of the centroid estimation of a particle/dot. The analysis is performed using an established theoretical framework called the Cramer-Rao Lower Bound (CRLB), a concept borrowed from the field of parameter estimation [12].

## 1.1 Cramer-Rao Lower Bound (CRLB)

In any experiment, the recorded measurement is a combination of the signal, which is the deterministic aspect of the measurement based on a physical model, and stochastic noise. Given the measurement, one would like to use a model for the measurement process to calculate a parameter of interest, where the particular method/algorithm used to calculate the parameter is called an estimator and the result obtained is the estimate. Based on the choice of the measurement and estimator, the estimates can have a bias (a systematic deviation from the true result) and a variance (due to the presence of noise), where a higher variance implies a lower measurement precision. The CRLB represents the lowest possible variance (or the highest possible precision) that can be achieved in the unbiased estimation of a parameter from a noise- and resolution-limited measurement. In the case of a biased estimation, the CRLB provides the lower bound on the random component of the error. It is a useful tool to study the scaling of error with respect to parameters in an experimental setup [12], [13].

Consider a signal/measurement **f**, which is composed of a known model signal **m** defined by a parameter $a$, and a noise component **n** such that $f_k = m_k + n_k$, $\forall k$, where $k$ represents the index of the temporal/spatial location at which the measurement is acquired. Further, let the signal be defined by a joint Probability Density Function (PDF) $p(\mathbf{f}, a)$. Then the CRLB for the estimate of $a$ is defined as: [13]

$$\sigma_a^2 = -\frac{1}{E\left[\frac{\partial^2 \ln p(\mathbf{f}, a)}{\partial a^2}\right]}. \qquad (1)$$

If the model is defined by a vector of parameters $\mathbf{a}$, then the CRLB is defined as the inverse of the diagonal element of the Fisher Information Matrix $J_{ij}$ which is defined as

$$J_{ij} = -E\left[\frac{\partial^2 \ln p(\mathbf{f}, \mathbf{a})}{\partial a_i \partial a_j}\right] \qquad (2)$$

with the CRLB for a parameter $a_i$ being given by $\sigma_{a_i}^2 = (J_{ii})^{-1}$ [13].

In the present context, the intensity of the particle/dot recorded on the camera is the *measurement*, contaminated by the thermal noise in the camera, and the sub-pixel fitting procedure is the *estimator*, to obtain the particle/dot centroid which is the *estimate*. We further assume that the only random process in the measurement chain is the thermal noise added to the sensor. In experiments, the true sub-pixel location of the particle/dot can also be a random variable, but this factor will be ignored in the current analysis and the *true centroid* of the particle/dot is assumed to be a constant value.

The CRLB has been derived for 2D PIV/PTV measurements in the past by Wernet and Pline [14], and Westerweel [15]. In both previous analyses, the CRLB was derived for locating the centroid of a Gaussian particle image discretely sampled on a CCD sensor in the presence of noise, but the two approaches differed in their assumptions about the probability density function (PDF) of the noise and thus provide different but complementary results. Wernet and Pline [14] considered the case of low illumination intensity where the noise is dominated by the photon count per pixel, which follows a Poisson distribution. Under this assumption, the CRLB was found to be linearly proportional to the particle diameter and inversely proportional to the photon count. Westerweel [15] considered the case of more recent PIV experiments with high-energy pulsed lasers, where the noise is primarily governed by the thermal noise in the CCD sensor, as well as the resolution and digitization noise, all of which are normally distributed. This assumption resulted in the lower bound being proportional to the square of the particle diameter and inversely proportional to the pixel pitch and illumination intensity.

Our work extends the analysis of Elsinga et. al. [5]–[7] using the theoretical framework of Westerweel [15] to characterize the effect of density/refractive-index gradients on a volumetric PTV/BOS experimental measurement. The analysis will be confined to the CRLB of the 2D position error, as the propagation of the 2D position error through the measurement chain of a volumetric PTV setup accounting for the calibration and reconstruction procedures is the subject of ongoing research. It will be seen that the effect of density/refractive-index gradients is to

*increase the lower bound* (or decrease the precision) of the centroid estimation process, thus decreasing the precision of the overall measurement. The increase in the CRLB is due to non-linearities in the density gradient field which result in blurring of the particle/dot image.

In the following sections, we first construct a comprehensive model for the particle/dot pattern image by considering the propagation of light rays through a medium with density/refractive-index gradients in a volumetric setup at an arbitrary viewing direction. We formulate the image model in terms of the experimental parameters: (1) particle depth, (2) viewing direction, (3) aperture f-number, and (4) image noise. We then construct the Fisher information matrix for the case of normally distributed noise due to the camera sensor and derive the CRLB. Finally, we compare the model predictions to ray-tracing simulations and demonstrate the uncertainty amplification using experimental BOS images.

## 2 Theory

### 2.1 Image Model

The overall imaging process can be represented using a transfer function approach as shown in Figure 1.. Each particle/dot acts as a point source of several light rays which travel through the density gradient field and the optical train and form an image on the camera sensor. The image due to a single light ray can be represented by the convolution of a Dirac delta function centered at the location of the geometric image $\vec{X}_r$ and the point spread function of the optical system $\tau(\vec{X})$. The collective image of the particle/dot formed by all light rays is given by,

$$I(\vec{X}) = \tau(\vec{X}) * \sum_{r=1}^{N_R} I_{0,r} \delta(\vec{X} - \vec{X}_r) \qquad . \tag{3}$$

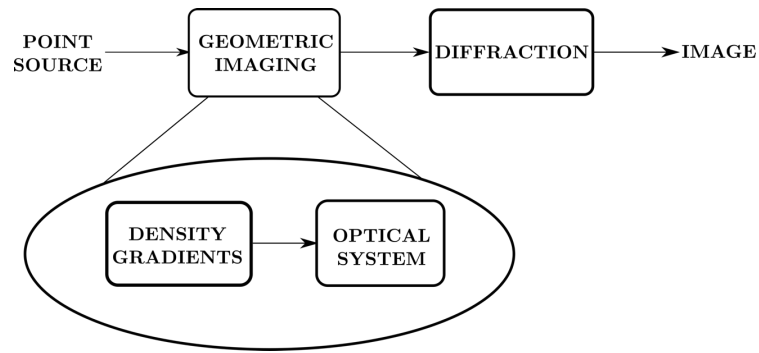

**Figure 1.** Illustration of the imaging process using a transfer function model.

The final location of the light ray $\vec{X}_r$ depends on the density gradients and the optical layout. For simplicity, we will first consider a head-on viewing configuration in the presence of density gradients, and subsequently generalize it to a case with an arbitrary viewing direction, and then model the effect of the aperture.

### 2.1.1 Head-on viewing and effect of particle depth

For the optical setup shown in Figure 2., the final location of a light ray $\vec{X}_r$ originating from a particle with initial conditions of $\vec{x}_r, \vec{\theta}_r$, can be expressed as

$$\vec{X}_r = M\vec{x}_r + \Delta\vec{X}_r \qquad (4)$$

where the first term is due to magnification of the imaging system, and the second term is the apparent displacement caused by the density gradients. We further use lower case symbols for co-ordinates in the object space and upper case symbols for co-ordinates in the image space on the camera.

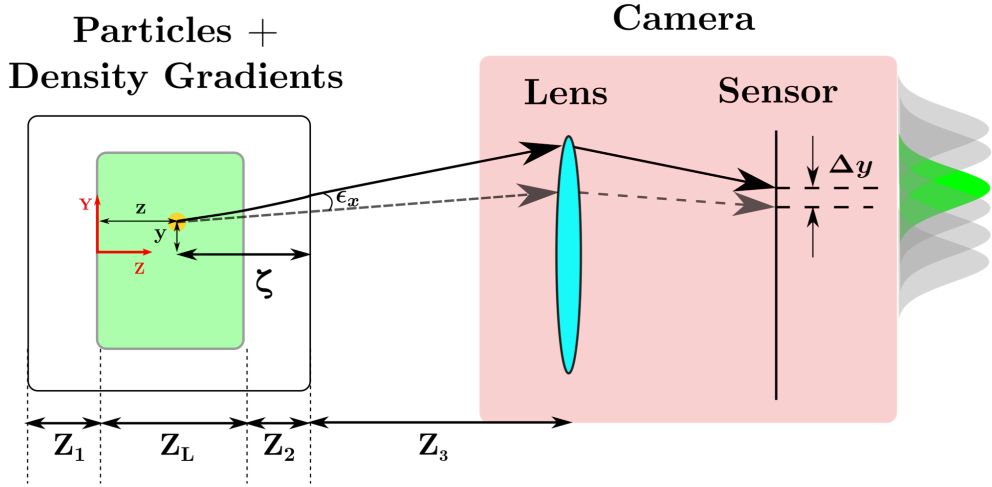

**Figure 2.** Schematic showing the imaging of a particle located in a three-dimensional measurement volume in the presence of density gradients.

The magnification $M$ is a function of the object distance and the focal length of the camera lens $f$, as given by,

$$M = \left(1 - \frac{Z_3 + \zeta}{f}\right)^{-1} \qquad (5)$$

and the apparent displacement of the light ray as it traverses a density gradient field is given by [2],

$$\Delta\vec{X}_r = \frac{MK}{n_0}\int_{z_i}^{z_f} \nabla\rho \, dz \approx \frac{MK}{n_0}\zeta^2(\nabla\rho)_{avg,r} \qquad (6)$$

where the integral is replaced by a path averaged value of the density gradient $(\nabla \rho)_{avg,r}$ experienced by each light ray. In the subsequent analysis, the subscript $avg$ will be dropped, with the understanding that the gradient field is always the path averaged value.

We have thereby modeled the effect of the particle depth in the measurement volume, $\zeta$, on the imaging process. While the magnification of the particle decreases with increasing depth, the depth appears as a quadratic term in the apparent displacement equation, and therefore, particles located deeper in the measurement volume will experience a larger apparent displacement of rays due to density gradients.

Finally, the point spread function of the optical system $\tau(\vec{X})$ is the intensity field created by Fraunhofer diffraction due to a circular aperture and is given by the Airy function. Typically, this is approximated by a Gaussian profile,

$$\tau(\vec{X}) = I_0 \exp\left(-\frac{|\vec{X}|^2}{2\eta^2}\right) \tag{7}$$

where $I_0$ is the peak intensity and $\eta$ is related to the diffraction diameter and is given by $\eta = \frac{1}{4}\frac{\sqrt{2}}{\pi}f_\#(1+M)\lambda$, where $\lambda$ is the wavelength of light [16]. Therefore, the model for the image of a single particle formed by all the light rays originating from the particle is given by,

$$I(\vec{X}) = \sum_{r=1}^{N_R} I_{0,r} \exp\left[-\frac{\left|\vec{X} - \left(M\,\vec{x}_r + \frac{MK}{n_0}\zeta^2 \nabla\rho|_r\right)\right|^2}{2\eta^2}\right]. \tag{8}$$

Here $I_{0,r}$ is the peak image intensity, and is defined as $I_{0,r} = \frac{\alpha_r}{2\pi\eta^2}$, where $\alpha_r$ is the total image exposure due to a single light ray.

### 2.1.2 Generalization to an arbitrary viewing direction

The effect of viewing direction on the imaging can be modeled using a rotation matrix to specify the camera orientation in three-dimensional space, and by expressing the relationship between the object and image space using a linear transformation [17],

$$\vec{X} = R\vec{x} + \vec{T} \tag{9}$$

Here $R = R_\phi R_\theta R_\psi$ is the rotation matrix and is a function of the pitch ($\phi$), yaw ($\theta$), and roll ($\psi$) angles of the camera, and $\vec{T}$ is the translation, which is a function of the object distance. The viewing configuration is illustrated in Figure 3..

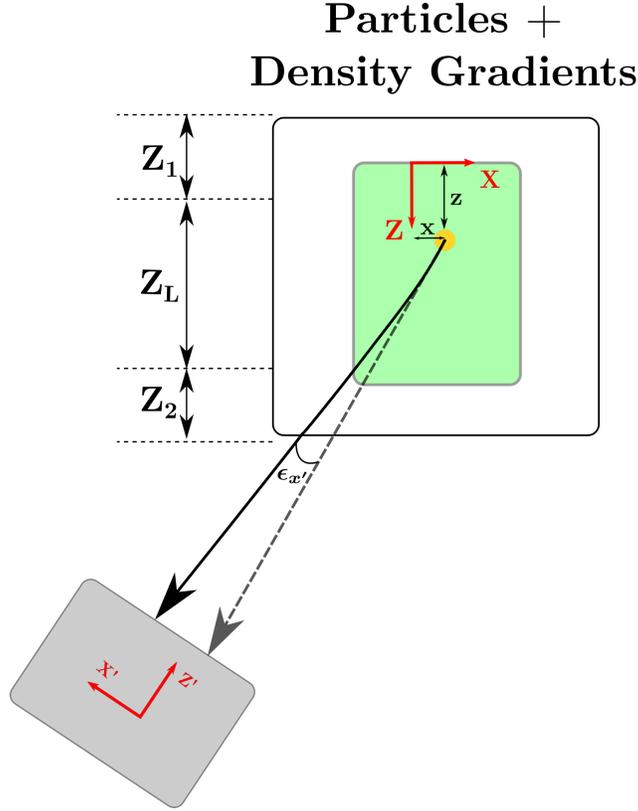

**Figure 3.** Illustration of the imaging layout for an arbitrary viewing direction.

The linear transformation can also be used to relate the density gradients in the two co-ordinate systems, and the equation for the apparent displacement can then be expressed in index notation as,

$$\Delta X_{r,j} = \frac{MK}{n_0} \zeta_l \zeta_l \frac{1}{R_{jk}} \nabla_k \rho \tag{10}$$

where $j, k$ and $l$ are tensor indices. Thus, we see that the apparent displacement is now a function of the angle between the viewing direction of the camera and the orientation of the density gradients in object space. Also, the case with the head-on viewing discussed previously becomes a special case when one of the camera angles is 90º and the other angles are zero.

### 2.1.3  *Effective image model and the role of the f-number*

To simplify further analysis, we rewrite Equation (8) to represent the intensity field due to an "effective Gaussian" with a centroid located at $(X_0, Y_0)$ and diameter $(4\eta_{0,x}, 4\eta_{0,y})$, under the assumption that the intensity field formed by several Gaussian distributions will also be a Gaussian distribution,

$$I_{eff}(\vec{X}) = I_0 \exp\left[-\left\{\frac{(X-X_0)^2}{2\eta_{0,x}^2} + \frac{(Y-Y_0)^2}{2\eta_{0,y}^2}\right\}\right] \quad . \tag{11}$$

The effective centroid and diameters can be estimated by equating the moments of the Gaussian distributions expressed in Equations (8) and (11). The main assumption that enables this simplification is that the viewing angle $\Delta\theta_0$ subtended by the particle on the lens is small, and therefore the density gradient field experienced by any arbitrary light ray can be expressed using a Taylor series expansion about the angular bisector of the ray cone, as illustrated in Figure 4. The viewing angle is a function of the object distance and the f-number, and for volumetric PTV/BOS applications, the f-number is generally high to provide a large depth of field. For volumetric PTV experiments, large f-numbers are required to keep a thick laser sheet in focus, and in BOS experiments, they are required to keep both the dot target and the density gradient field in focus. These requirements result in a very small viewing angle, on the order of 1-2 degrees, which makes the small angle assumption reasonable.

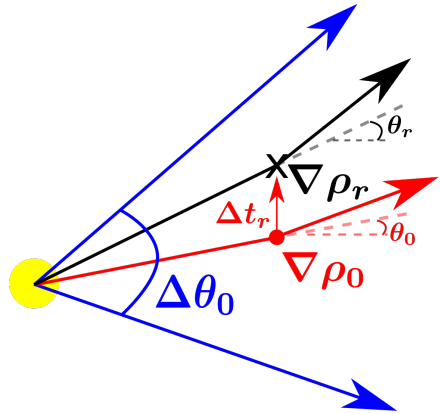

**Figure 4.** Illustration of the Taylor series approximation for the density gradient field showing the ray cone (bounded by the blue rays), the angular bisector (red) and an arbitrary light ray (black). The dashed lines are the light ray trajectories in the absence of a density gradient field.

The effective centroid is described as the first moment of the intensity distribution of the image and is given by,

$$\vec{X}_0 = \frac{\int_{-\infty}^{+\infty}\int_{-\infty}^{+\infty} \vec{X} I(\vec{X}) dX\, dY}{\int_{-\infty}^{+\infty}\int_{-\infty}^{+\infty} I(\vec{X}) dX\, dY} \quad . \tag{12}$$

The above equation will be simplified for the x component in the following analysis, and the procedure for the y component is identical. We have, for $\vec{X}_0 = X_0\hat{\imath} + Y_0\hat{\jmath}$,

$$X_0 = \frac{\int_{-\infty}^{+\infty}\int_{-\infty}^{+\infty} X \sum_{r=1}^{N_R} I_{0,r} \exp\left[-\frac{|\vec{X}-\vec{X}_r|^2}{2\eta^2}\right] dX\, dY}{\int_{-\infty}^{+\infty}\int_{-\infty}^{+\infty} \sum_{r=1}^{N_R} I_{0,r} \exp\left[-\frac{|\vec{X}-\vec{X}_r|^2}{2\eta^2}\right] dX\, dY}$$

$$= \frac{\sum_{r=1}^{N_R} I_{0,r} \int_{-\infty}^{+\infty}\int_{-\infty}^{+\infty} X \exp\left[-\frac{|\vec{X}-\vec{X}_r|^2}{2\eta^2}\right] dX\, dY}{\sum_{r=1}^{N_R} I_{0,r} \int_{-\infty}^{+\infty}\int_{-\infty}^{+\infty} \exp\left[-\frac{|\vec{X}-\vec{X}_r|^2}{2\eta^2}\right] dX\, dY}$$

$$= \frac{\sum_{r=1}^{N_R} I_{0,r} \int_{-\infty}^{+\infty} X \exp\left[-\frac{(X-X_r)^2}{2\eta^2}\right] dX \int_{-\infty}^{+\infty} \exp\left[-\frac{(Y-Y_r)^2}{2\eta^2}\right] dY}{\sum_{r=1}^{N_R} I_{0,r} \int_{-\infty}^{+\infty} \exp\left[-\frac{(X-X_r)^2}{2\eta^2}\right] dX \int_{-\infty}^{+\infty} \exp\left[-\frac{(Y-Y_r)^2}{2\eta^2}\right] dY}$$

$$= \frac{\sum_{r=1}^{N_R} I_{0,r}(\sqrt{2\pi}X_r\eta)(\sqrt{2\pi}\eta)}{\sum_{r=1}^{N_R} I_{0,r}(\sqrt{2\pi}\eta)(\sqrt{2\pi}\eta)} \tag{13}$$

where we have used the result for Gaussian integrals, that $\int_{-\infty}^{+\infty} \exp\left[-\frac{(x-a)^2}{2b^2}\right] dx = \sqrt{2\pi}b$ and $\int_{-\infty}^{+\infty} x \exp\left[-\frac{(x-a)^2}{2b^2}\right] dx = \sqrt{2\pi}ab$. Further, under the assumption of a small angle of the ray cone, we expect that $I_{0,r} \approx I_0 = constant$ for all light rays emerging from a particle. Note however, that $I_0$ can still be different for a particle viewed from different cameras, as well as particles at different locations in the laser sheet due to Mie scattering. If the cone angle is very small, then all rays in the cone will travel approximately in the same direction and the variation of the Mie scattering coefficients over the solid angle of the ray cone for the same camera can be neglected.

Further, for a small cone angle, it is also expected that all light rays travel approximately the same distance from the particle to the edge of the density gradient field ($\zeta_r \approx \zeta = constant$). Based on these assumptions, we obtain that

$$X_0 = \frac{1}{N_R} \sum_{r=1}^{N_R} X_r$$

$$= \frac{1}{N_R} \sum_{r=1}^{N_R} \left(M x_0 + \frac{MK}{n_0} \zeta_r^2 \frac{\partial \rho}{\partial x}\bigg|_r\right)$$

$$= M x_0 + \frac{MK\zeta^2}{n_0 N_R} \sum_{r=1}^{N_R} \frac{\partial \rho}{\partial x}\bigg|_r \quad . \tag{14}$$

The second term in the above equation represents the *average deflection* of all light rays originating from the particle, and it can be simplified further by using a Taylor series expansion of the density gradient field about the angular bisector of the ray cone. Consider a straight line connecting the particle, the center of the lens and a point on the camera sensor, which represents the case of pinhole imaging. Let the intersection point of this line with the end-plane of the density gradient field be $\vec{t}_0 = \vec{x}_0 + \zeta\vec{\theta}_0$ and the density gradient at this intersection point be defined as $\nabla\rho|_{t_0}$. Then the density gradient along the x-direction at an arbitrary intersection point $\vec{t}_r$ can be expressed to first order as

$$\left.\frac{\partial\rho}{\partial x}\right|_{\vec{t}_r} = \left.\frac{\partial\rho}{\partial x}\right|_{\vec{t}_0} + \left.\frac{\partial^2\rho}{\partial x^2}\right|_{\vec{t}_0}(t_{r,x} - t_{0,x}) + \left.\frac{\partial^2\rho}{\partial x \partial y}\right|_{\vec{t}_0}(t_{r,y} - t_{0,y}) + O(|t_r - t_0|^2) . \quad (15)$$

The parameter $(\vec{t}_r - \vec{t}_0)$ can be expressed in terms of the optical setup as

$$\begin{aligned}\vec{t}_r - \vec{t}_0 &= (\vec{x}_0 + \zeta\vec{\theta}_r) - (\vec{x}_0 + \zeta\vec{\theta}_0) \\ &= \zeta(\vec{\theta}_r - \vec{\theta}_0)\end{aligned} \quad (16)$$

with $\vec{\theta}_r$ being the angle of propagation of a light ray, and $\vec{\theta}_0 = \frac{1}{N_R}\sum_{r=1}^{N_R}\vec{\theta}_r$ being the angle between the bisector of the cone and the horizontal.

Using Equations (15) and (16) in Equation (14) and simplifying, we obtain for the effective centroid along x,

$$\begin{aligned}X_0 &= Mx_0 + \frac{MK\zeta^2}{n_0 N_R}\sum_{r=1}^{N_R}\left(\left.\frac{\partial\rho}{\partial x}\right|_0 + \left.\frac{\partial^2\rho}{\partial x^2}\right|_0 \zeta(\theta_{r,x} - \theta_{0,x}) + \left.\frac{\partial^2\rho}{\partial x \partial y}\right|_0 \zeta(\theta_{r,y} - \theta_{0,y})\right) \\ &= Mx_0 + \frac{MK\zeta^2}{n_0}\left(\left.\frac{\partial\rho}{\partial x}\right|_0 + \left.\frac{\partial^2\rho}{\partial x^2}\right|_0 \zeta\frac{1}{N_R}\sum_{r=1}^{N_R}(\theta_{r,x} - \theta_{0,x}) + \left.\frac{\partial^2\rho}{\partial x \partial y}\right|_0 \zeta\frac{1}{N_R}\sum_{r=1}^{N_R}(\theta_{r,y} - \theta_{0,y})\right) \\ &= Mx_0 + \frac{MK\zeta^2}{n_0}\left.\frac{\partial\rho}{\partial x}\right|_0 \quad .\end{aligned} \quad (17)$$

where the summations evaluate to zero due to the definition of $\theta_{0,x}$ and $\theta_{0,y}$ as the angles of the bisectors. The y component of the centroid can be evaluated in a similar manner, and the 2D centroid of the particle/dot image is given by,

$$\vec{X}_0 = M\vec{x}_0 + \frac{|M|K\zeta^2}{N_R n_0}\nabla\rho|_0 \quad (18)$$

where the first term on the right hand side is the image location for a particle without the density gradients and the second term is the average displacement of light rays due to the density gradient field, with $\nabla\rho|_0$ representing the depth-averaged density gradient experienced by the angular bisector of the ray cone.

The effective particle/dot diameter can be computed by equating the second moment of the Gaussian distributions. Again, just considering the x-component, the standard deviation (a measure of the diameter) is given by:

$$\eta_{0,x}^2 = \frac{\int_{-\infty}^{+\infty}\int_{-\infty}^{+\infty}(X-X_0)^2 I(\vec{X})dX\,dY}{\int_{-\infty}^{+\infty}\int_{-\infty}^{+\infty} I(\vec{X})dX\,dY}$$

$$= \frac{\sum_{r=1}^{N_R} I_{0,r}\int_{-\infty}^{+\infty}(X-X_0)^2 \exp\left[-\frac{(X-X_r)^2}{2\eta^2}\right]dX \int_{-\infty}^{+\infty}\exp\left[-\frac{(Y-Y_r)^2}{2\eta^2}\right]dY}{\sum_{r=1}^{N_R} I_{0,r}\int_{-\infty}^{+\infty}\exp\left[-\frac{(X-X_r)^2}{2\eta^2}\right]dX \int_{-\infty}^{+\infty}\exp\left[-\frac{(Y-Y_r)^2}{2\eta^2}\right]dY}$$

$$= \frac{\sum_{r=1}^{N_R} I_{0,r}\left(\sqrt{2\pi}\eta(\eta^2 + (X_r-X_0)^2)\right)\left(\sqrt{2\pi}\eta\right)}{\sum_{r=1}^{N_R} I_{0,r}\left(\sqrt{2\pi}\eta\right)\left(\sqrt{2\pi}\eta\right)}$$

$$= \eta^2 + \frac{1}{N_R}\sum_{r=1}^{N_R}(X_r - X_0)^2 \qquad (19)$$

again using the previously mentioned results for the Gaussian integrals in addition to a new result, that $\int_{-\infty}^{+\infty}(x-c)^2 \exp\left[-\frac{(x-a)^2}{2b^2}\right]dx = \sqrt{2\pi}b\{(a-c)^2 + b^2\}$.

The second term in the summation can again be simplified using the Taylor series expansion as follows,

$$X_r - X_0 = Mx_0 + \frac{MK\zeta^2}{n_0}\frac{\partial\rho}{\partial x}\bigg|_r - \left(Mx_0 + \frac{MK\zeta^2}{n_0}\frac{\partial\rho}{\partial x}\bigg|_0\right)$$

$$= \frac{MK\zeta^2}{n_0}\left(\frac{\partial\rho}{\partial x}\bigg|_r - \frac{\partial\rho}{\partial x}\bigg|_0\right)$$

$$= \frac{MK\zeta^2}{n_0}\left(\frac{\partial^2\rho}{\partial x^2}\bigg|_0 (t_{r,x} - t_{0,x}) + \frac{\partial^2\rho}{\partial x\partial y}\bigg|_0 (t_{r,y} - t_{0,y})\right)$$

$$= \frac{MK\zeta^2}{n_0}\left(\frac{\partial^2\rho}{\partial x^2}\bigg|_0 \zeta(\theta_{r,x} - \theta_{0,x}) + \frac{\partial^2\rho}{\partial x\partial y}\bigg|_0 \zeta(\theta_{r,y} - \theta_{0,y})\right). \qquad (20)$$

The summation then becomes,

$$\frac{1}{N_R}\sum_{r=1}^{N_R}(X_r - X_0)^2 = \left(\frac{MK\zeta^2}{n_0}\right)^2 \zeta^2 \frac{1}{N_R}\sum_{r=1}^{N_R}\left(\left.\frac{\partial^2 \rho}{\partial x^2}\right|_0 (\theta_{r,x} - \theta_{0,x}) + \left.\frac{\partial^2 \rho}{\partial x \partial y}\right|_0 (\theta_{r,y} - \theta_{0,y})\right)^2$$

$$= \left(\frac{MK\zeta^2}{n_0}\right)^2 \zeta^2 \left(\left(\left.\frac{\partial^2 \rho}{\partial x^2}\right|_0\right)^2 \frac{1}{N_R}\sum_{r=1}^{N_R}(\theta_{r,x} - \theta_{0,x})^2\right.$$

$$+ \left(\left.\frac{\partial^2 \rho}{\partial x \partial y}\right|_0\right)^2 \frac{1}{N_R}\sum_{r=1}^{N_R}(\theta_{r,y} - \theta_{0,y})^2$$

$$\left. + \left.\frac{\partial^2 \rho}{\partial x^2}\right|_0 \left.\frac{\partial^2 \rho}{\partial x \partial y}\right|_0 \frac{2}{N_R}\sum_{r=1}^{N_R}(\theta_{r,x} - \theta_{0,x})(\theta_{r,y} - \theta_{0,y})\right) \quad (21)$$

The summations of the angles can be simplified further by modeling the angular distribution of light rays as a uniform random variable, where the angle of propagation of any given light ray is randomly distributed within the total angle of the ray cone $\Delta\theta_{0,x}$. That is,

$$P_{\Theta_x}(\theta_{r,x}) = \begin{cases} \frac{1}{\Delta\theta_{0,x}} & \theta_{0,x} - \frac{\Delta\theta_{0,x}}{2} \leq \theta_{r,x} \leq \theta_{0,x} + \frac{\Delta\theta_{0,x}}{2} \\ 0 & \text{otherwise} \end{cases} \quad (22)$$

with a similar expression for the random variable $\Theta_y$ corresponding the distribution of the angle of propagation along y, $\theta_{r,y}$, which would depend on the component of the cone angle along y, $\Delta\theta_{0,y}$. Except for highly astigmatic viewing configurations, the components of the cone angle along the two directions will be equal ($\Delta\theta_{0,x} = \Delta\theta_{0,y} = \Delta\theta_0$). By further assuming that the random variables along the two components $\Theta_x$ and $\Theta_y$ are independent, it can be shown that

a) $\frac{1}{N_R}\sum_{r=1}^{N_R}(\theta_{r,x} - \theta_{0,x})^2 = \sigma_{\Theta_x} = \frac{\Delta\theta_0^2}{12}$

b) $\frac{1}{N_R}\sum_{r=1}^{N_R}(\theta_{r,y} - \theta_{0,y})^2 = \sigma_{\Theta_y} = \frac{\Delta\theta_0^2}{12}$

c) $\frac{1}{N_R}\sum_{r=1}^{N_R}(\theta_{r,x} - \theta_{0,x})(\theta_{r,y} - \theta_{0,y}) = 0$

and the summation simplifies to,

$$\frac{1}{N_R}\sum_{r=1}^{N_R}(X_r - X_0)^2 = \frac{1}{12}\left(\frac{MK\zeta^2}{n_0}\right)^2 \zeta^2 \Delta\theta_0^2 \left(\left(\left.\frac{\partial^2 \rho}{\partial x^2}\right|_0\right)^2 + \left(\left.\frac{\partial^2 \rho}{\partial x \partial y}\right|_0\right)^2\right). \quad (23)$$

The angle of the ray cone $\Delta\theta_0$ can be expressed in terms of the parameters of the optical setup as

$$\Delta\theta_0 = \frac{1}{f_\#}\left(1 + \frac{1}{M}\right)^{-1} - \frac{K\zeta}{2n_0}\left(\left.\frac{\partial \rho}{\partial x}\right|_0 + \left.\frac{\partial \rho}{\partial y}\right|_0\right) \quad (24)$$

where the second term accounts for the mean deflection of the ray cone due to density gradients. The standard deviation ( = ¼ diameter) along the x direction is then given by,

$$\eta_{0,x}^2 = \eta^2 + \frac{1}{12}\left(\frac{MK\zeta^2}{n_0}\right)^2 \zeta^2 \Delta\theta_0^2 \left(\left(\frac{\partial^2 \rho}{\partial x^2}\bigg|_0\right)^2 + \left(\frac{\partial^2 \rho}{\partial x \partial y}\bigg|_0\right)^2\right) \quad (25)$$

with a similar analysis to yield the standard deviation along the y direction,

$$\eta_{0,y}^2 = \eta^2 + \frac{1}{12}\left(\frac{MK\zeta^2}{n_0}\right)^2 \zeta^2 \Delta\theta_0^2 \left(\left(\frac{\partial^2 \rho}{\partial x \partial y}\bigg|_0\right)^2 + \left(\frac{\partial^2 \rho}{\partial y^2}\bigg|_0\right)^2\right) \quad (26)$$

Equations (25) and (26) show that non-linearities in the density gradient field increase the effective diameter of the dot leading to *blurring*, consistent with earlier observations by Elsinga et. al. [5] for 2D PIV from experimental data. Also, using the effective image model, we have thereby modeled the effect of the f-number, which is the third parameter of interest. It is to be noted that in the case of BOS experiments, the term $\frac{MK\zeta^2}{n_0} = \frac{MKZ_D \Delta z}{n_0}$, where $Z_D$ is the distance between the dot pattern and the density gradient field and $\Delta z$ is the depth of the gradient field.

Finally, the peak image intensity of the effective model can be expressed in terms of the image exposure $\alpha_0$ defined as,

$$\begin{aligned}
\alpha_0 &= \int_{-\infty}^{+\infty}\int_{-\infty}^{+\infty} I(X,Y) dX\, dY \\
&= I_0 \int_{-\infty}^{+\infty} \exp\left[-\frac{(X-X_0)^2}{2\eta_{0,x}^2}\right] \int_{-\infty}^{+\infty} \exp\left[-\frac{(Y-Y_0)^2}{2\eta_{0,y}^2}\right] dX\, dY \\
&= 2\pi I_0 \eta_{0,x} \eta_{0,y}
\end{aligned} \quad (27)$$

Further, since the exposure for the effective model should be equal to that formed by all the individual light rays,

$$\begin{aligned}
\alpha_0 &= \int_{-\infty}^{+\infty}\int_{-\infty}^{+\infty} I(X,Y) dX\, dY \\
&= \int_{-\infty}^{+\infty}\int_{-\infty}^{+\infty} \sum_{r=1}^{N_R} I_r(X,Y)\, dX\, dY \\
&= \sum_{r=1}^{N_R} \int_{-\infty}^{+\infty}\int_{-\infty}^{+\infty} I_r(X,Y)\, dX\, dY \\
&= \sum_{r=1}^{N_R} \alpha_r,
\end{aligned} \quad (28)$$

Thus, the peak intensity for the effective image model is given by

$$I_0 = \frac{\alpha_0}{2\pi\eta_{0,x}\eta_{0,y}} \quad , \tag{29}$$

thereby completing the formulation of the effective image model.

The final image of the particle/dot sampled on a discrete set of pixels and with a finite number of gray levels is given by,

$$g_{kl} = \gamma d_r^2 I(X_k, Y_l) \quad . \tag{30}$$

Here $k, l$ are the pixel indices along the $X, Y$ directions respectively, $\gamma$ is the pixel to gray level conversion factor, and $d_r$ is the pixel pitch. In addition, all CCD/CMOS sensors have some amount of noise added to the signal due to thermal noise and finite number of gray levels. This will be modeled in the next section when constructing the Fisher Information Matrix.

## 2.2 Noise Model and Fisher Information

The final image of the particle/dot recorded on the sensor is the sum of the sampled intensity profile $g_{kl}$ as given by Equation (30) with some additive noise $\hat{n}_{kl}$:

$$\hat{g}_{kl} = \gamma d_r^2 I(X_k, Y_l) + \hat{n}_{kl} \quad . \tag{31}$$

Following Westerweel [15], the fluctuations due to thermal noise and finite number of gray levels are assumed to be normally distributed, signal-independent and uncorrelated, with a standard deviation of $\sigma_n$. The joint pdf of the measurement then becomes

$$p(\hat{g}, \boldsymbol{a}) = \left(\frac{1}{2\pi\sigma_n^2}\right)^{\frac{MN}{2}} \exp\left[-\frac{1}{2\sigma_n^2}\sum_{k=0}^{M}\sum_{l=0}^{N}(\hat{g}_{kl} - g_{kl})^2\right] \tag{32}$$

where $\boldsymbol{a}$ is the parameter vector and $M, N$ are the number of pixels along $X$ and $Y$, respectively.

The Fisher information matrix defined earlier represents the total amount of information available about the particle/dot from its intensity profile that can be used to estimate its centroid. For the present scenario, the Fisher Information available to estimate the $X$ component of the centroid becomes [13], [15]:

$$
\begin{aligned}
J_{X_0 X_0} &= \frac{1}{\sigma_n^2} \sum_{k=0}^{M} \sum_{l=0}^{N} \left(\frac{\partial g_{kl}}{\partial X_0}\right)^2 \\
&= \left(\frac{\gamma I_0 d_r^2}{\sigma_n \eta_{0,x}^2}\right)^2 \sum_{k=0}^{M} (X_k - X_0)^2 \exp\left[-\frac{(X_k - X_0)^2}{2\eta_{0,x}^2}\right] \sum_{l=0}^{N} \exp\left[-\frac{(Y_l - Y_0)^2}{2\eta_{0,y}^2}\right] \\
&= \left(\frac{\gamma I_0 d_r}{\sigma_n \eta_{0,x}^2}\right)^2 \int_{-\infty}^{+\infty} (X_k - X_0)^2 \exp\left[-\frac{(X_k - X_0)^2}{2\eta_{0,x}^2}\right] dX_k \int_{-\infty}^{+\infty} \exp\left[-\frac{(Y_l - Y_0)^2}{2\eta_{0,y}^2}\right] dY_l \\
&= \left(\frac{\gamma I_0 d_r}{\sigma_n}\right)^2 \frac{\pi}{2} \frac{\eta_{0,y}}{\eta_{0,x}} \ .
\end{aligned}
\tag{33}
$$

where the summations are converted into integrals under the assumption that the extent of the dot is small compared to the size of the whole camera sensor.

## 2.3 Cramer-Rao Lower Bound

The Cramer-Rao lower bound for the variance is defined as the inverse of the diagonal elements of the Fisher Information matrix, and therefore the standard deviation for the estimation of the centroid is given by

$$
\begin{aligned}
\sigma_{X_0 X_0} &= \sqrt{\frac{1}{J_{X_0 X_0}}} \\
&= \sqrt{\frac{2}{\pi} \frac{\eta_{0,x}}{\eta_{0,y}}} \frac{\sigma_n}{\gamma I_0 d_r} \ .
\end{aligned}
\tag{34}
$$

Further, the peak intensity $I_0$ can be related to the particle/dot diameter using Equation (29) to obtain

$$
\sigma_{X_0 X_0} = \frac{2\sqrt{2\pi} \sigma_n}{\gamma \alpha_0 d_r} \eta_{0,x}^{\frac{3}{2}} \eta_{0,y}^{\frac{1}{2}} \ .
\tag{35}
$$

In the absence of blurring, $\eta_{0,x} = \eta_{0,y} = \eta$ and the result simplifies to

$$
\sigma_{X_0 X_0} = \frac{2\sqrt{2\pi} \sigma_n}{\gamma \alpha_0 d_r} \eta^2
\tag{36}
$$

thereby recovering a result similar to Westerweel [15], that the lower bound increases with the square of the particle/dot diameter. Further, since it was shown previously that the diameter increases due to nonlinearities in the density gradient field, the effect of the density gradients is to *increase* the variance in the measurement of the centroid. For a linear density gradient field, there is a uniform translation of the particle/dot and the CRLB is then identical to the result obtained by Westerweel for PIV.

In summary, the Cramer-Rao lower bound associated with the estimation of a 2D particle/dot centroid from the image in the presence of density gradients and thermal noise in a volumetric setup, is given by Equation (35), and is a function of:

1) Exposure ($\alpha_0$) which is a function of the viewing direction in the case of Mie scattering
2) Noise level ($\sigma_n$)
3) Diffraction Diameter ($4\eta$)
4) Magnification ($M$)
5) Particle Depth ($\zeta$)
6) Non-linearities in the density field and viewing direction ($\nabla^2 \rho, R_{ij}$)
7) Camera Aperture ($f_\#$)

In the case of a 2D planar PIV experiment, the magnification, viewing direction, and particle depth, are constants for a given experiment, and we obtain the result that the variance is only a function of the aperture setting and the nonlinearities in the density field, consistent with the experimental result obtained by Elsinga et. al. [5]

In the case of a BOS experiment, the depth variable represents the distance between the dot target and the density gradient field ($\zeta = Z_D$), and the blurring is only present for the image with the density gradients, and is absent for the reference image. Further, the parameter $MZ_D$ is defined as the *sensitivity* of a BOS setup, with a larger value being considered better for resolving small scale features in the flow field [18]. However, since it also increases the variance and lowers the precision of the measurement in the presence of blur, an optimal trade-off between sensitivity and precision should be considered when designing an experiment. Further, increasing $f_\#$ tends to reduce the blur and increase the measurement precision, and a large $f_\#$ also helps to keep both the dot pattern and the density gradient field in focus. Therefore, small aperture settings must be favored when designing BOS experiments, which generally requires the use of high power illumination.

## 2.4 Uncertainty Amplification Ratio

The direct application of Equation (35) to calculate the CRLB for a given experiment is limited by the accuracy in the estimation of the image noise level which can vary across the field of view. In order to isolate the uncertainty amplification due to density gradients on the CRLB, an amplification ratio metric is proposed for BOS experiments that removes the effect of properties that are common to both the reference and gradient images. The amplification ratio, $AR$, is defined as the ratio of the CRLBs for the same dot in the reference and the gradient images, and it can be shown that this ratio is purely a function of the ratios of the peak intensities and ratios of the dot diameters,

$$AR_X = \frac{\sigma_{X_0 X_{0_{grad}}}}{\sigma_{X_0 X_{0_{ref}}}} = \sqrt{\frac{\eta_{0,x_{grad}}}{\eta_{0,x_{ref}}} \frac{\eta_{0,y_{ref}}}{\eta_{0,y_{grad}}} \frac{I_{0_{ref}}}{I_{0_{grad}}}}$$

$$AR_Y = \frac{\sigma_{Y_0 Y_{0_{grad}}}}{\sigma_{Y_0 Y_{0_{ref}}}} = \sqrt{\frac{\eta_{0,y_{grad}}}{\eta_{0,y_{ref}}} \frac{\eta_{0,x_{ref}}}{\eta_{0,x_{grad}}} \frac{I_{0_{ref}}}{I_{0_{grad}}}} \quad . \tag{37}$$

In addition, the proposed amplification ratio metric can also be used to report position uncertainties in tracking-based processing for BOS, as these processing methods are shown to significantly improve the accuracy, precision and spatial resolution [19]. First, the position uncertainty for the reference image can be calculated by recording several (e.g. 1000) images of the dot pattern without any flow and estimating the standard deviation of the dot centroids from the subpixel fits. Then, the position uncertainty for a given dot in the gradient image can be calculated by multiplying the position uncertainty for the corresponding dot in the reference image and the amplification ratio. The uncertainty in the density field can then be obtained by propagating the displacement uncertainties through the BOS measurement chain [20], [21].

## 3 Comparison of the model with simulations

The theoretical result for the CRLB obtained in the previous section was then compared with ray tracing simulations. We have developed a ray tracing-based synthetic image generation methodology for PIV/BOS experiments in variable density environments, for the purpose of simulating general optical setups as well as for error and uncertainty analysis [22]. Using this methodology, synthetic images of particles under diffraction limited imaging and viewed through a known density field were generated by randomly varying four parameters (1) non-linearities in the density field $\left(\frac{\partial^2 \rho}{\partial x^2}, \frac{\partial^2 \rho}{\partial x \partial y}, \frac{\partial^2 \rho}{\partial y^2}\right)$, (2) particle depth, (3) f-number and (4) viewing direction. The parameters were drawn from a uniform distribution and the ranges of values considered were selected based on typical experimental setups, as listed in Table 1. In particular, the range of values of the second gradient of density was taken from experimental results of Elsinga et. al [23] for a Prandtl-Meyer expansion fan at a 11° corner. A total of 1000 images were generated using this parameter range.

**Table 1.** Range of parameters used to generate particle images in the Monte-Carlo simulations.

| Parameter | Range |
|---|---|
| $\frac{\partial^2 \rho}{\partial x^2}, \frac{\partial^2 \rho}{\partial x \partial y}, \frac{\partial^2 \rho}{\partial y^2}$ | 0 to 5000 kg/m^5 |
| Particle Depth | -2 mm to +2 mm |
| F-number | 8, 11, 16 |
| Viewing direction (x and y) | -10 to 10 deg. |

For each realization of the Monte-Carlo simulation, the particle image was corrupted with noise modeled as a random variable drawn from a zero-mean normal distribution. For the simulations reported in this paper, the standard deviation of the normal distribution was set to be 5% of the peak intensity of the particle image.

The particle/dot centroids in the noisy image were located using a Least Squares Gaussian (LSG) subpixel fitting scheme, which involves fitting a Gaussian curve to the intensity map of a particle, under the assumption that the diffraction limited image can be approximated by a Gaussian distribution [16], [24]. The Gaussian curve is described by five parameters: (1) X location of the centroid $(X_0)$, (2) Y location of the centroid $(Y_0)$, (3) peak intensity $(I_0)$, (4) diameter/standard deviation along x, $(\eta_{0,x})$ and (5) standard deviation along y $(\eta_{0,x})$ where the particle diameter is defined as four times of the standard deviation, consistent with PIV literature [24], [25]. In this method, the parameters are obtained by minimizing the residual between the predicted intensity from the fit and the actual intensity of the pixels using a non-linear least squares method. The least squares method is initialized by the three-point Gaussian (TPG) fit, which calculates the centroid, peak intensity and standard deviation of a Gaussian curve by fitting it to the intensity values recorded at three points/pixels on the dot intensity map [26]. The three points are taken to be the pixel with maximum intensity and one pixel on either side of the maximum. The centroid estimates are then compared with the ground truth obtained from the final positions of the light rays traced in the simulation to compute the position error. For each case of the Monte-Carlo simulation, the noise addition and centroid estimation procedure was performed 1000 times to calculate the variance of the position error.

The LSG fit is the Maximum Likelihood Estimate (MLE) for the centroid estimation in the case of normally distributed image noise. Since the MLE approaches the CRLB in the limit of large number of observations, it is expected that the variance of the centroid estimates from the LSG method would approach the result for the CRLB derived in the previous section [13].

It can be shown that the LSG is the MLE for the centroid estimation problem as follows. For a probability density function defined as $p_\theta(x)$ where $x$ is the signal, and $\theta$ is the parameter governing the pdf, the likelihood function is defined as

$$L(\theta|x) = p_\theta(x) \quad . \tag{38}$$

Typically, the likelihood function is replaced by the log likelihood given by

$$l(\theta|x) = \ln L(\theta|x) \quad . \tag{39}$$

The maximum likelihood estimate for the parameter $\theta$ given a set of measurements $x$ obeying a known pdf $p_\theta(x)$ is defined as the value that maximizes the likelihood function for the given set of measurements. Since the log function is monotonic, maximizing the log likelihood function is equivalent to maximizing the likelihood function.

$$\hat{\theta} \in \underset{\theta \in \Theta}{\operatorname{argmax}} \, l(\theta|x) \quad . \tag{40}$$

The example we consider here is the Gaussian image of a dot on a camera sensor corrupted with zero-mean noise that is normally distributed with standard deviation $\sigma_n$. It was shown earlier that the joint PDF of the image intensity over a M x N pixel region is given by

$$p(\hat{g}, a) = \left(\frac{1}{2\pi\sigma_n^2}\right)^{\frac{MN}{2}} \exp\left[-\frac{1}{2\sigma_n^2}\sum_{k=0}^{M}\sum_{l=0}^{N}(\hat{g}_{kl} - g_{kl})^2\right]. \tag{41}$$

Therefore, the log likelihood function for the parameter $a$ given an intensity measurement $\hat{g}$ becomes

$$l(a|\hat{g}) = -\frac{MN}{2}\ln 2\pi\sigma_n^2 - \frac{1}{2\sigma_n^2}\sum_{k=0}^{M}\sum_{l=0}^{N}(\hat{g}_{kl} - g_{kl})^2 \tag{42}$$

and the MLE becomes

$$\hat{a} \in \underset{\theta \in \Theta}{\operatorname{argmax}} \, l(a|\hat{g})$$
$$\in \underset{\theta \in \Theta}{\operatorname{argmin}} \sum_{k=0}^{M}\sum_{l=0}^{N}(\hat{g}_{kl} - g_{kl})^2 \tag{43}$$

which is a Least Squares solution. Since $g_{kl}$ is modelled to be a Gaussian, the MLE is a Least Squares Gaussian fit.

The results obtained from the Monte-Carlo simulations are shown in Figure 5 where the model prediction for the standard deviation of the position error (CRLB) are compared to the results from the simulations. We show that while there is a reasonable match between the model predictions and the simulations, the model under-predicts the variance from the simulation by about 0.005 pix. This mismatch could be attributed to two possible reasons. First, the MLE is expected to converge to the CRLB in the limit of an infinite number of observations, so it is possible that the finite number of observations could result in a slight over-estimation of the MLE. Second, the image model derived in the paper is a simplified description of the full non-linear trajectory of light rays propagating through the density gradient medium, and hence it is expected to under-predict the blurring of the particle images. Therefore, a combination of the over-estimation of the standard deviation from the simulation due to the finite number of observations and an under-prediction due to the simplifications in the model derivations is most likely the reason for the disagreement between the results. However, even with the large number of simplifications, the model is able to reasonably predict the trend of the position estimation variance, and hence can provide useful scaling rules in the design of experiments.

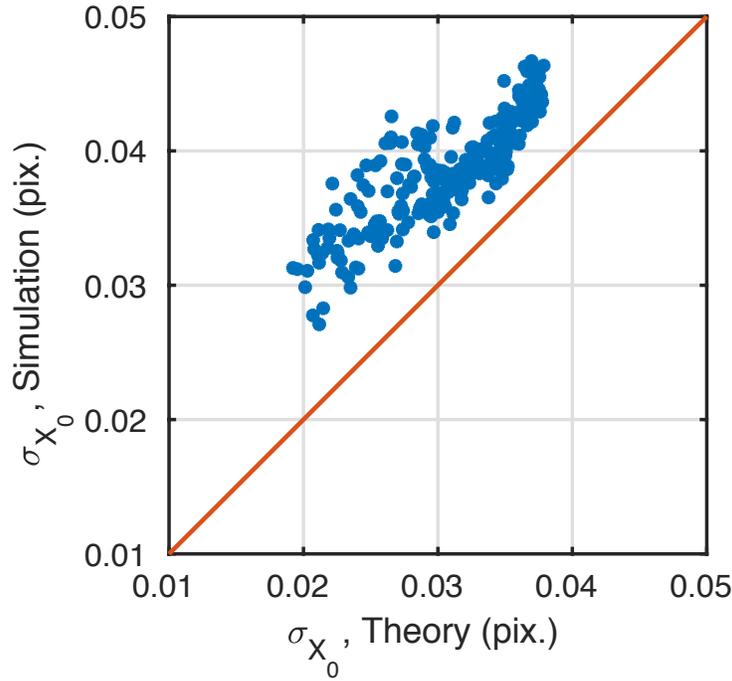

**Figure 5.** Comparison of standard deviation of the position error between model predictions (CRLB) given by Equation (35) and results from Monte-Carlo simulations

## 4 Demonstration with Experimental Images

The proposed CRLB estimation methodology and the uncertainty amplification is demonstrated on experimental BOS images of flow induced by a nanosecond spark discharge reported by Singh et. al. [27], [28]. The spark discharge leads to rapid heating of the gas in the electrode gap resulting in a complex three-dimensional flow field. The BOS measurements were performed by recording images of a target dot pattern containing a regular grid of dots in the presence of the spark induced flow field. The images with the flow can be compared to a reference image recorded prior to the discharge to estimate the displacement field and to measure the blurring of the dots in the presence of density gradients. More details of the experimental setup can be found in Singh et. al. [27], [28].

For the experimental images, the Uncertainty Amplification Ratio $AR$ defined in Section 2.4 was calculated for all the identified dots in both the reference and gradient images using an elliptical Least Square Gaussian fit, and the magnitude of the amplification ratio ($= \sqrt{AR_x^2 + AR_y^2}$) is shown in Figure 6.. Also shown are the displacement field which corresponds to the projected density gradients as given by Equation (6) and a histogram of the position uncertainty magnitude for the gradient image calculated using the uncertainty quantification methodology outlined in Section 2.4. The dot identification and displacement estimation is performed using a newly developed dot tracking methodology for BOS measurements that has been shown to significantly improve the accuracy, precision and spatial resolution of the displacement estimation process [11]. The method utilizes prior knowledge of the positions and sizes of the dots on the target to improve the identification and centroid estimation process.

The figures show that regions corresponding to the displacement gradients are coincident to regions with large values of the ratio metric. A value of the ratio metric greater than 1 implies that the CRLB for the gradient image will be higher than the reference image in this region. While the regions greater than 1 mostly occur in regions with large second gradients of the density (first derivative of displacement), there are still some stray values in regions without density gradients, that are most likely due to intensity fluctuations between the reference and gradient images from the Xenon arc lamp light source used for the experiments.

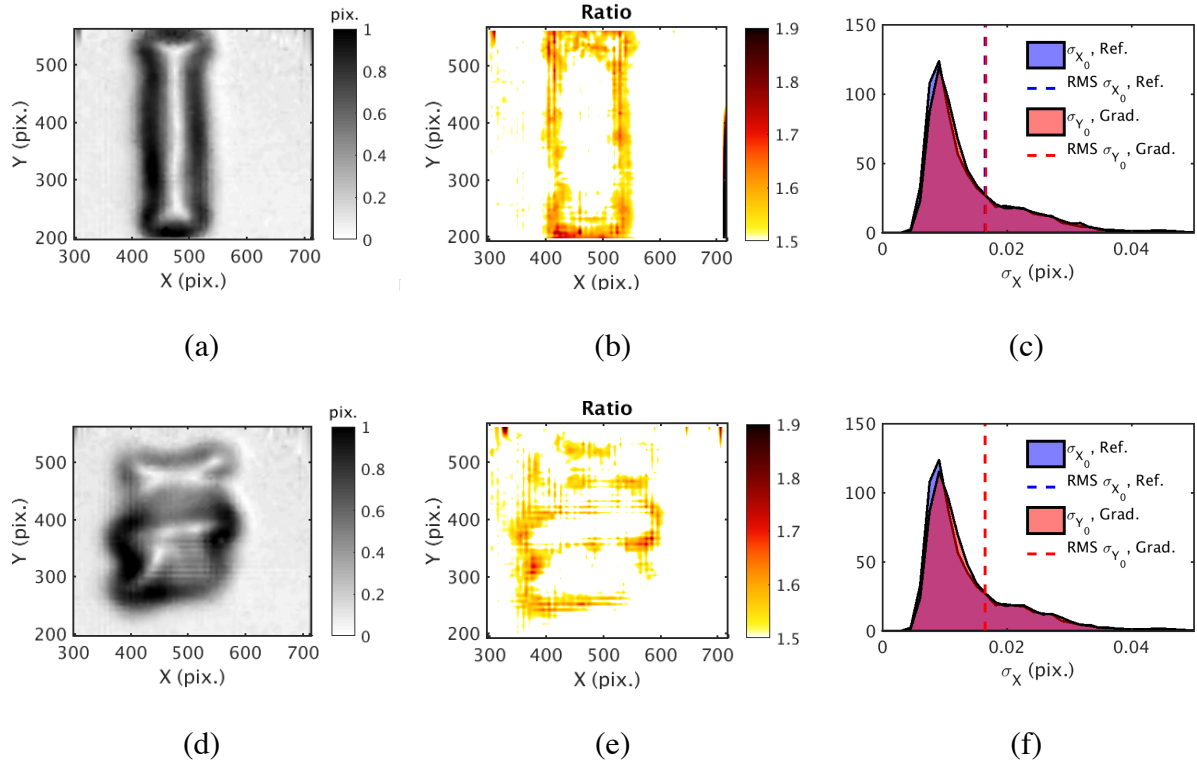

**Figure 6.** Snapshots of the displacements (a), (d), amplification ratios (b), (e), and histogram of position uncertainties for the gradient image (c), (f) for two time instants of the spark induced flow field.

## 5 Conclusions

The effect of density/refractive-index gradients on the precision of volumetric PTV/BOS experiments was theoretically analyzed using the Cramer-Rao lower bound of the 2D centroid estimation process. To perform the analysis, we derived a model for the diffraction limited image of a particle/dot viewed through a non-linear density gradient field under the assumption of a small ray cone angle which is expected to be reasonable given the requirement of ensuring a thick laser sheet in volumetric PTV experiments and for keeping both the target and the flow field in focus for BOS experiments. Under the further assumption that the effective model of a particle/dot imaged through density gradients can be described by a Gaussian profile, we showed that the effective centroid is given by the original centroid of the image in addition to a shift corresponding to the average deflection of light rays, and that the effective diameter is a root mean squared sum of the diffraction diameter and a blurring due to the second derivatives of the density field. Further, we also showed that the relative angle between the viewing direction and the density gradient field can also have an effect on the particle image. As a result of the increase in the diameter due to blurring, the effect of density gradients is to *increase* the Cramer-Rao lower bound and to lower the measurement precision in the centroid estimation process. For BOS, it was also seen that the ratio of the CRLBs of the dots in the reference and gradient images, termed the Uncertainty Amplification Ratio ($AR$), is a function of the ratio of their diameters and the peak intensities. Based on this ratio, a methodology was proposed to report position uncertainties for tracking-based BOS measurements.

The Cramer-Rao lower bound predicted by the imaging model was then compared with the position estimation variance obtained from ray tracing simulations for diffraction limited imaging of a particle in a thick laser sheet viewed through a density gradient field from different directions. The rendered particle images were corrupted with Gaussian noise and the centroid was estimated using a Least Square Gaussian fit, and the corresponding position error was calculated. The variance of this position error from 1000 trials was then compared to the CRLB for a particular parameter set. This procedure was repeated for 1000 random combinations of the experimental parameters. The results show that the model, given the various assumptions and simplifications, is able to predict the position estimation variance from the ray tracing simulations, though there is a slight under-prediction. The implications of the model of the CRLB were also demonstrated with experimental BOS images of flow induced by a nanosecond spark discharge. Analysis of the images showed that the CRLB for the gradient image is amplified with respect to the reference image particularly in regions of strong density gradients.

The CRLB for the 2D centroid estimation process derived in this work can also be propagated through the measurement chain of a volumetric PTV setup accounting for uncertainties introduced in the calibration and reconstruction procedures to elucidate the effect of distortions due to density/refractive-index gradients on the 3D centroid estimation process [29]. Further, the possible utility of the uncertainty amplification factor for uncertainty quantification for correlation-based PIV/BOS similar to the peak-to-peak ratio and signal to noise ratio metrics [30]–[32] is an avenue for further work.

## 6 Acknowledgement

This material is based upon work supported by the U.S. Department of Energy, Office of Science, Office of Fusion Energy Sciences under Award Number DE-SC0018156. Bhavini Singh is acknowledged for help with the spark discharge experiment. Jiacheng Zhang, Javad Eshragi and Adib Ahmadzadegan are acknowledged for reviewing the manuscript.